\documentclass[runningheads]{llncs}
\usepackage[T1]{fontenc}
\usepackage{microtype}
\usepackage{graphicx}
\usepackage{subfigure}
\usepackage{booktabs} 
\usepackage{hyperref}

\usepackage{algorithm}
\usepackage{algorithmic} 

\usepackage{amsmath}
\usepackage{amssymb}
\usepackage{mathtools}

\usepackage{enumitem}
\usepackage{tabularx}
\usepackage{multirow}
\usepackage{makecell}
\usepackage{colortbl}
\usepackage{placeins}
\usepackage{mathrsfs}
\usepackage{xcolor}

\begin{document}
\title{From First-Order to Second-Order Rationality: Advancing Game Convergence with Dynamic Weighted Fictitious Play}
\titlerunning{From First-Order to Second-Order Rationality}
%
\author{
    Qi Ju\inst{1,2} \and
    Falin Hei\inst{1,2} \and
    Yuxuan Liu\inst{1,2}\and
    Zhemei Fang\inst{1,2}\and
    Yunfeng Luo\inst{1,2}
}
%
%
\institute{Huazhong University of Science and Technology \and
National Key Laboratory of Science and Technology on Multispectral Information Processing\\
\email{juqi@hust.edu.cn}
}

\maketitle              
    \begin{abstract}
        Constructing effective algorithms to converge to Nash Equilibrium (NE) is a important problem in algorithmic game theory.
        Prior research generally posits that the upper bound on the convergence rate for games is $O\left(T^{-1/2}\right)$.
        This paper introduces a novel perspective,
        positing that the key to accelerating convergence in game theory is “rationality”.
        Based on this concept,
        we propose a Dynamic Weighted Fictitious Play (DW-FP) algorithm.
        We demonstrate that this algorithm can converge to a NE and exhibits a convergence rate of $O(T^{-1})$ in experimental evaluations.
        \keywords{Game Theory  \and Rationality \and Fictitious Play \and Dynamic Weight.}
    \end{abstract}

    \footnotetext{This work has been accepted by PRICAI 2024.}
    \section{Introduction}\label{sec:introduction}
    An important issue in algorithmic game theory is solving Nash Equilibrium (NE).
Traditionally,
solving for an exact NE is an extremely complex non-linear programming problem.
Currently, it is more common to use an iterative approach to gradually approach an NE.
Notably, 
the Regret Minimization algorithm (RM), its variants and Fictitious Play (FP)~\cite{brown1951iterative,fudenberg1998theory},
have achieved significant success in games like poker~\cite{bowling2015heads,brown2019superhuman} and Starcraft~\cite{vinyals2019grandmaster}.

Variants of RM primarily fall into two categories:
those extending RM's applicability,
such as CFR~\cite{zinkevich2007regret},
Monte Carlo-RM/CFR~\cite{lanctot2009monte},
RM/CFR pruning~\cite{brown2015regret};
and those enhancing RM's rate of convergence,
like RM/CFR+~\cite{tammelin2014solving},
discount RM/CFR~\cite{brown2019solving},
and Greedy RM~\cite{zhang2022equilibrium}.
Although these algorithms perform well in practice,
theoretically,
these algorithms can only guarantee that the distance between the strategy obtained after $T$ iterations and the NE strategy is $O\left(T^{-1/2}\right)$
(in the following article, this distance is called the convergence rate).

The previous iterative solution algorithms inherit from online optimization ones.
These methods only consider choosing the current optimal strategy from their own perspective.
We've improved by breaking individual decision-making in training.
All agents disclose their actions in training (only during training; it's undisclosed in actual games).
Each agent adjusts its iterative weight considering everyone's strategies.
We call decision based only on one's own situation first-order rationality;
considering oneself and opponents is second-order rationality.
Applying this to FP gives a new DW-FP algorithm.
We prove its convergence to NE and faster convergence speed.
Our code can be found at \href{https://github.com/Zealoter/Dynamic-Weighted-Fictitious-Play}{GitHub}.
    \section{Notation and Preliminaries}\label{sec:notation-and-preliminaries}
    \subsection{Game Theory}\label{subsec:game-theory}
\subsubsection{Normal-Form Games}
In normal-form games, the set of players is denoted by $\mathcal{N}=\{1,2,\dots\}$.
For each player $i$,
there is a finite set of legal actions $\mathcal{A}^i=\{a_1,a_2,\dots\}$.
The mixed strategy set $\Sigma ^i\in \mathbb{R}^{|\mathcal{A}|}$ (where $|\cdot|$ represents the number of elements in the set) is a probability distribution over $\mathcal{A}^i$.
If a strategy assigns probability $1$ to a single action and $0$ to others, it is termed pure strategy.
We directly use $a$ to represent the corresponding pure strategy.
A strategy profile $\sigma=\times_{i \in \mathcal{N} }\sigma^i$ is formed by the strategies of all players,
and $\sigma^{-i}$ represents the strategy profile of all players except $i$.
The payoff function $u^i:\Sigma\rightarrow \mathbb{R}$ is finite,
and $u^i(a,\sigma^{-i})$ denotes the payoff for player $i$ when taking action $a$ while other players follow the strategy profile $\sigma^{-i}$.

\subsubsection{NE and $\epsilon$-NE}
The best response (BR) set for player $i$ against the strategy profile $\sigma^{-i}$ of the opponents is defined as
\begin{equation}
    b^i(\sigma^{-i})={\arg\max}_{a^{*}\in \mathcal{A}^i}u^i(a^{*},\sigma^{-i}).
    \label{eq:201}
\end{equation}
Player $i^\prime$s exploitability $\epsilon^i$ for a strategy profile $\sigma$ is defined as
\begin{equation}
    \epsilon^i=u^i(b^i(\sigma^{-i}),\sigma^{-i})-u^i(\sigma),
    \label{eq:202}
\end{equation}
the overall exploitability $\epsilon$ of all players is defined as $\epsilon=\frac{1}{|\mathcal{N} |}{\sum}_{i\in \mathcal{N} }\epsilon^i$,
when $\epsilon=0$,
the strategy profile $\sigma$ is a NE,
otherwise it is a $\epsilon$-NE.
Directly solve an accurate NE in games is a PPAD problem~\cite{papadimitriou1994complexity}.
In practice,
it is more common to use an iterative method to get a $\epsilon$-NE .
During iterations,
if the exploitability satisfies $\epsilon \propto T^{-1/2}$,
the convergence rate of the algorithm is $O\left(T^{-1/2}\right)$.

\subsection{Regret Minimization and Greedy Regret Minimization}\label{subsec:regret-and-greedy-regret-minimization}
\begin{algorithm}[h]
    \caption{Greedy RM}
    \label{alg:greedy_rm}
    \begin{algorithmic}[1]
        \STATE Random initialization $\sigma_{t=1}$. Set $\bar{\sigma}_{t=1}^i=\boldsymbol{0}$, $\bar{R}_{t=0}^{i} = \boldsymbol{0}$, $W^i_{t=0}=0$ for all $i\in \mathcal{N}$.
        \FOR{$t=1,2,\dots$}
            \STATE $R^i_{t}(a)=u^{i}\left(a, \sigma^{-i}_{t-1}\right)-u^{i}\left(\sigma_{t-1}\right)$
            \STATE $w^i_t=\min_{w^\prime}\Phi \left(\frac{W^i_{t-1}}{W^i_{t-1}+w^\prime}\bar{R}_{t-1}^{i}+\frac{w^\prime}{W^i_{t-1}+w^\prime} R^i_{t}\right) $
            \STATE $\alpha_t^i=\frac{w^i_t}{W^i_{t-1}+w^i_t}$
            \STATE $\bar{R}_{t}^{i}=(1-\alpha_t^i)\bar{R}_{t-1}^{i}+\alpha_t^i R^i_{t}$
            \STATE $\bar{R}_{t}^{i,+}(a)=\max\left(\bar{R}_{t}^{i}(a),0 \right)$
            \IF{$\sum_{a \in \mathcal{A}^i} \bar{R}_{t}^{i,+}(a)>0$}
                \STATE $\sigma_{t+1}^{i}(a)=\frac{\bar{R}_{t}^{i,+}(a)}{\sum_{a \in \mathcal{A}^i} \bar{R}_{t}^{i,+}(a)}$
            \ELSE
                \STATE $\sigma_{t+1}^{i}(a)= \frac{1}{|\mathcal{A}^i|}$
            \ENDIF

            \STATE $\bar\sigma_t=(1-\alpha_t^i)\bar{\sigma}_{t-1}+\alpha_t^i\sigma_{t}$
            \STATE $W^i_{t}=W^i_{t-1}+w^i_t$
        \ENDFOR
    \end{algorithmic}
\end{algorithm}

For any strategy sequence $\sigma_1 \ldots \sigma_T$ in the game,
the pseudo-code of Greedy RM as shown in ~\ref{alg:greedy_rm}.
Here, $R^i\left(a^{i\prime}\right)$ regret value, $\Phi$ is the potential function,
defined as
\begin{equation}
    \Phi({R}^{i})=\sum_{i\in \mathcal{N}}\sum_{a\in\mathcal{A}^i}{R}^{i,+}(a).
    \label{eq:203}
\end{equation}

In the original RM,
the weight at each iteration is $w_t=1$.
Greedy RM shows a convergence speed of $O(T^{-1})$ in practical problems.
However,
only $O\left(T^{-1/2}\right)$ can be proven at present like original RM~\cite{cesa2006prediction}.

\subsection{Fictitious Play}\label{subsec:fictitious-play-and-its-variants}
\begin{algorithm}[h]
    \caption{Q-Value Based FP}
    \label{alg:fp}
    \begin{algorithmic}[1]
        \STATE Random initialization $\sigma_{t=1}$. Set $\bar{\sigma}_{t=1}^i=\boldsymbol{0}$ $\bar{Q}_{t=0}^{i} = \boldsymbol{0}$ for all $i\in\mathcal{N}$.
        \FOR{$t=1,2,\dots$}
            \STATE $\bar{Q}_{t}^{i}(a) = \frac{t-1}{t} \bar{Q}^i_{t-1}(a) + \frac{1}{t} u(a, \sigma_t^{-i})$
            \STATE $\sigma^i_{t+1} = \arg\max_{a \in \mathcal{A}^i} \bar{Q}^i_t(a)$
            \STATE $\bar\sigma_t=\frac{t-1}{t}\bar{\sigma}_{t-1}+\frac{1}{t}\sigma_{t}$
        \ENDFOR
    \end{algorithmic}
\end{algorithm}

The pseudo-code of Greedy FP as shown in ~\ref{alg:fp}.
From the FP process, it can be found that the next strategy $\sigma_{t+1}^i$ is the action with the maximum Q-value $Q_t^i$.
If the index of the maximum Q-value does not change,
then $\sigma_{t+1}^i$ will not change.
Understanding this point will be of great help in understanding the idea of our algorithm.
    \section{Method}\label{sec:method}
    Just as athletes don't play full games in training and specialized training brings greater improvement,
this applies in game theory training too.
RM/FP agents assume opponents target them fully at every training iteration,
so they adopt the most favorable strategy in current iteration.
We call this agent-centered view for next iteration first-order rationality.
However, in game theory training, this assumption isn't necessary.
A cooperation mechanism can be introduced where agents disclose their next iteration strategies and adjust weights based on both their own and opponents' strategies,
making training more efficient(This is just like the specialized training adopted by athletes).
This view considering both one's own and opponents' strategies is called second-order rationality.

\subsection{Dynamic Weighted Fictitious Play}\label{subsec:dynamic-weighted-fp}
In FP,
as we noted in~\ref{subsec:fictitious-play-and-its-variants},
a strategy change occurs when there's a shift in the maximum of $\bar{Q}_t$.
Our task is to determine the number of iterations required for this shift in the maximum Q-value.
During training,
everyone discloses their strategies (only in training, not in actual games),
and then adjusts weights based on their own and opponents' disclosed strategies.
Solving dynamic weight can be seen as a pursuit problem.
Distance is $Q^{gap,i}(a)$,
speed is $S^i(a)$, so catch-up time is $w=Q^{gap}/S$.
According to $w$,
we can calculate how many iterations the current strategy $\sigma$ will last.
Skipping repetitive calculations simplifies the process and accelerates convergence.
The specific implementation of DW-FP is in Algorithm~\ref{alg:p301}.

\begin{algorithm}[h]
    \caption{DW-FP{\textbf{(Ours)}}}
    \label{alg:p301}
    \begin{algorithmic}[1]
        \STATE Random initialization $\sigma_{t=1}$. Set $\bar{\sigma}_{t=1}^i=\boldsymbol{0}$, $\bar{Q}_{t=0}^{i} = \boldsymbol{0}$, $W^i_{t=0}=0$ for all $i\in\mathcal{N}$.
        \FOR{$t=1,2,\dots$}
            \STATE $Q_t^{gap,i}(a)=W_{t-1}\max_{a^*\in \mathcal{A}^i}\left(\bar{Q}^i_{t-1}(a^*)-\bar{Q}^i_{t-1}(a) \right)$
            \STATE $S^i(a)=u^i(a,\sigma_t^{-i})-u^i\left(\sigma\right)$
            \IF{$S^i(a)>0$}
                \STATE $w^i_{t}(a)=Q_t^{gap,i}(a)/S^i(a) $
            \ELSE
                \STATE $w^i_{t}(a)=\inf$
            \ENDIF
            \STATE $w_{t}=\min_{i\in\mathcal{N}}\min_{a\in\mathcal{A} }w^i_{t}(a)$
            \STATE $\alpha_t^i=\frac{w_{t}}{W_{t-1}+w_{t}}$
            \STATE $\bar{Q}_{t}^{i}=(1-\alpha_t^i)\bar{Q}^i_{t-1}+\alpha_t^i u^i(a,\sigma_t^{-i})$
            \STATE $\sigma^i_{t+1}={\arg\max}_{a\in \mathcal{A}^i}\bar{Q}^i_t$
            \STATE $\bar{\sigma}_{t}^i(I)=(1-\alpha_t^i)\bar{\sigma}_{t-1}^i(I)+\alpha_t^i\sigma_{t}^i(I)$
            \STATE $W_{t}^i=W_{t-1}^i+w_{t}^i$
        \ENDFOR
    \end{algorithmic}
\end{algorithm}
    \section{Experimental Results}\label{sec:experiments}
    \begin{figure*}
    \centering
    \includegraphics[width=1.0\textwidth]{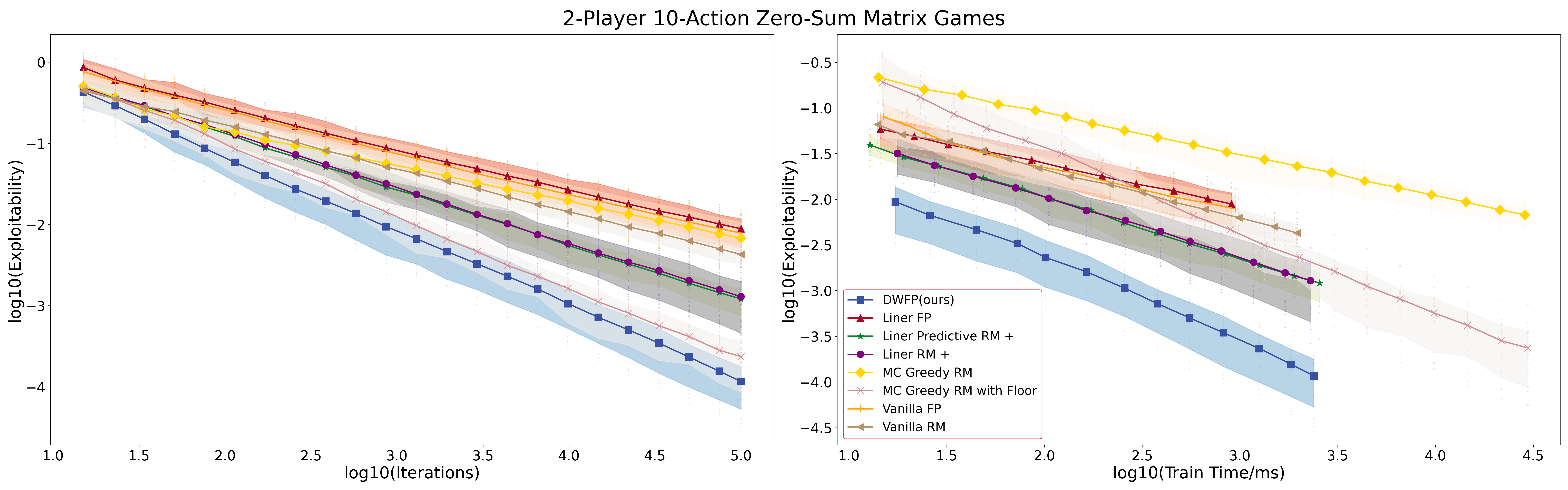}
    \caption{
        The figure illustrates the performance of various algorithms on 30 randomly generated two-player zero-sum games,
        with matrix payoff entries drawn from a standard normal Gaussian distribution $N(0,1)$.
        The shaded regions around each line represent the 95\% confidence interval.
        This setting will also be used in subsequent experiments.
    }
    \label{fig:normal_from}
\end{figure*}

We utilized algorithms such as FP, RM, RM+, and Greedy RM to compare with the proposed DW-FP in normal-form games (matrix games).
Notably, Greedy RM and RM+ have been recognized as particularly powerful in their respective game categories.
The selection of experiment parameters was guided by established best practices~\cite{brown2020equilibrium,zhang2022equilibrium}.
\subsection{Two-Player Zero-Sum Normal-Form Games}\label{subsec:two-player-zero-sum-normal-form-games}
Figure~\ref{fig:normal_from} shows that DW-FP has the fastest convergence rate,
outperforming RM+ significantly in both iteration numbers and time.
Furthermore, although the potential function for dynamic weights $w_t^i$ in Greedy RM can be approximated via bisection,
solving this optimization problem each time poses significant challenges.
Our experiments indicate that a DW-FP iteration takes only one-tenth the time of a Greedy RM iteration,
effectively highlighting DW-FP's advantages for engineering implementations.
\subsection{The Convergence Rate of DW-FP}\label{subsec:the-convergence-rate-of-dw-fp}
\begin{figure}
    \centering
    \includegraphics[width=0.6\textwidth]{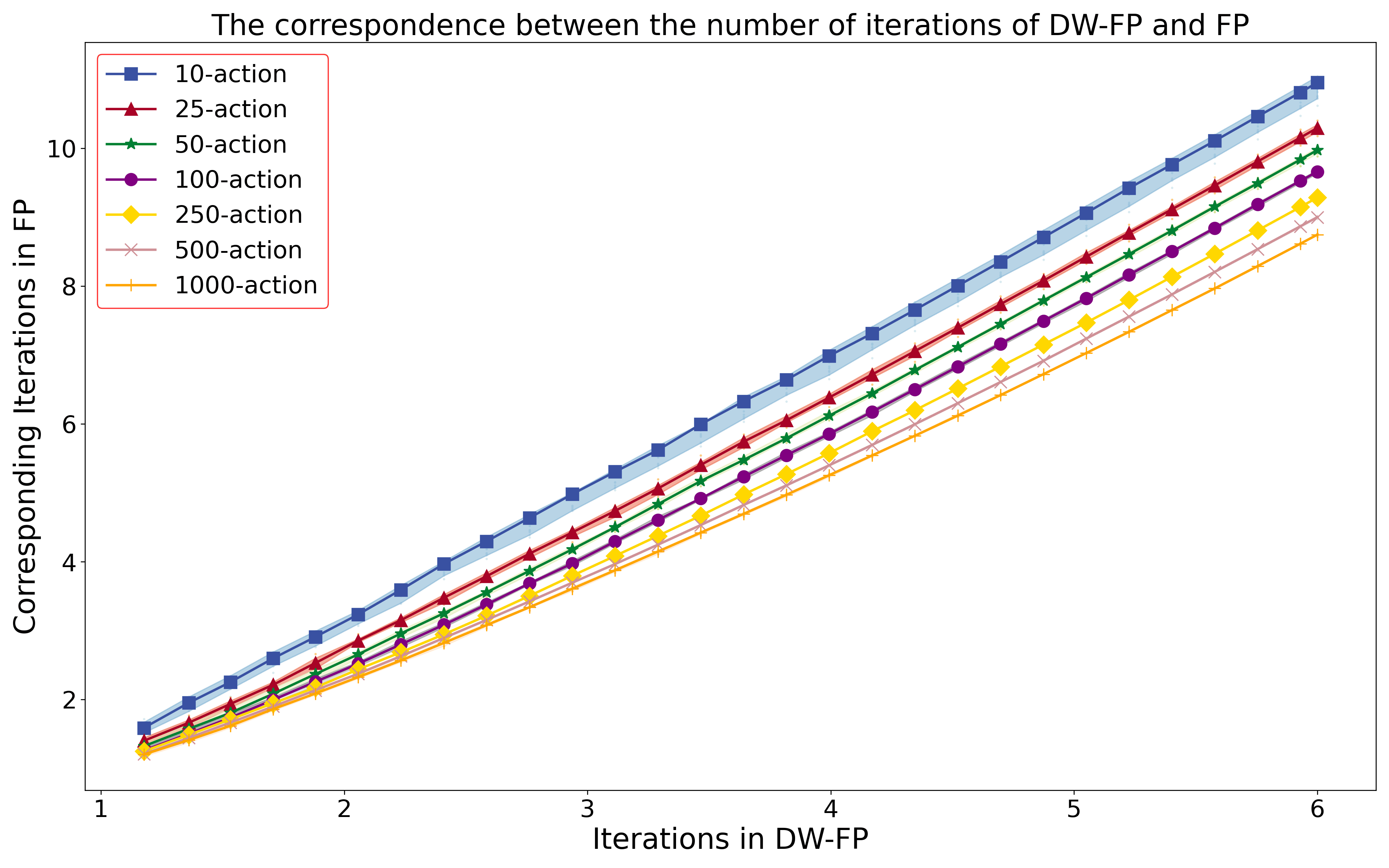}
    \caption{
        This figure indicates how many iterations of the original FP correspond to the iteration of DW-FP.
    }
    \label{fig:c_nor}
\end{figure}

DW-FP and FP are essentially the same,
except that DW-FP omits the iterative steps of FP through dynamic weights.
This raises a crucial question: How many original FP iterations are equivalent to a single iteration of DW-FP?
As illustrated in Figure~\ref{fig:c_nor},
the iteration for DW-FP is roughly the square of the iteration number for original FP.
Considering that the convergence speed of the original FP is $O\left(T^{-1/2}\right)$,
then the convergence speed of DW-FP may be $O(T^{-1})$.
    \section{Conclusions and Future Work}\label{sec:conclusions-and-future-work}
    This study presents the Dynamic Weighted Fictitious Play (DW-FP) algorithm.
It showcases the integration of “second-order rationality” to boost convergence rates in two player zero-sum games.
Further refinements could focus on expanding DW-FP's application to scenarios like extensive-form games.

    \appendix

\end{document}